# Microstructured fiber links for THz communications and their fabrication using infinity printing

GUOFU XU*, KATHIRVEL NALLAPPAN*, YANG CAO, AND
MAKSIM SKOROBOGATIY**

*Department of Engineering Physics, École Polytechnique de Montréal, Montreal, Québec, H3T 1J4, Canada*
*\* These authors contributed equally to the paper*
*\*\*maksim.skorobogatiy@polymtl.ca*

**Abstract:** In this work, a novel infinity 3D printing technique is explored to fabricate continuous multi-meter-long low-loss near-zero dispersion suspended-core polypropylene fibers for application in terahertz (THz) communications. Particular attention is paid to process parameter optimization for 3D printing with low-loss polypropylene plastic. Three microstructured THz fibers were 3D printed using the standard and infinity 3D printers, and an in-depth theoretical and experimental comparison between the fibers were carried out. Transmission losses (by power) of 4.79 dB/m, 17.34 dB/m and 11.13 dB/m are experimentally demonstrated for the three fibers operating at 128 GHz. Signal transmission with BER far below the forward error correction limit ($10^{-3}$) for the corresponding three fiber types of lengths of 2 m, 0.75 m and 1.6 m are observed, and an error-free transmission is realized at the bit rates up to 5.2 Gbps. THz imaging of the fiber near-field is used to visualize modal distributions and study optimal fiber excitation conditions. The ability of shielding the fundamental mode from the environment, mechanical robustness and ease of handling of thus developed effectively single-mode high optical performance fibers make them excellent candidates for upcoming fiber-assisted THz communications. Additionally, novel fused deposition modeling (FDM)-based infinity printing technique allows continuous fabrication of unlimited in length fibers of complex transverse geometries using advanced thermoplastic composites, which, in our opinion, is poised to become a key fabrication technique for advanced terahertz fiber manufacturing.

## 1. Introduction

The terahertz (THz) spectral range (0.1 THz-10 THz) has recently attracted much attention because of many potential applications in sensing [1, 2], imaging [3, 4], security [5, 6] and communications [7, 8]. In communications industry, in particular, the global IP traffic is increasing exponentially and is expected to reach 396 exabytes per month by 2022 [9]. Among the total traffic, more than 85% will be utilized by video, gaming and multimedia applications which requires reliable transmission of several tens of gigabytes per second per user. In this scenario, the wireless links for transmitting large data volumes is a particularly convenient modality for mobile end-users. According to Shannon theorem, for a given signal-to-noise ratio, the maximum capacity of the channel is limited by its bandwidth. In order to support higher bit rates, shifting the carrier frequency towards higher frequency range (THz spectral band and hence higher bandwidth) is one promising solution. Nevertheless, there are several challenges in realizing the free space wireless THz links. Firstly, due to self-diffraction of the electromagnetic waves, the free space path loss increases as a square of the distance between the transmitting and receiving antennas, even in absence of the atmospheric absorption. In turn, atmospheric absorption of the THz waves is strongly sensitive to the environmental conditions, which make such links unreliable for all-wather operation [10, 11]. Moreover, the signal interference effects can become pronounced in the ultra-dense THz wireless networks [12]. Additionally, due to higher frequency of THz waves compared to microwave radiation, they are more prone to scattering on various obstacles and structural imperfections, thus making non-line of sight THz communication challenging [13]. Finally, strong directionality of the THz beams demands tight alignment tolerances between the transmitting and receiving antennas for high quality THz wireless links [14]. In this respect, the THz waveguides/fibers can be an alternative solution to the medium-length (~10s of meters) THz communications links due to the fiber small footprint and flexibility, as well as ease of handling and installations even in complex geometrical environments [15]. Moreover, the THz fibers are immune to environmental variations, external electromagnetic interference and eavesdropping promising reliable and secure communications [16]. Ultimately, seamless integration of THz wireless links with THz fiber-assisted links can offer a reliable performance in the next generation communication systems. In such systems one can envision, for example, that a THz fiber network will be installed within the geometrically complex communication environment (such as a multistory building) to provide reliable points of THz wireless access within smaller and less complex communication environments (such as individual offices).

In order to realize efficient THz fiber-assisted communication links, the THz fibers must feature low transmission loss, low bending loss, low group velocity dispersion (GVD), high coupling efficiency, as well as good mechanical stability and low sensitivity to the environment variations[15]. The low loss guidance in THz fibers is generally achieved by adapting designs that allow significant fraction of THz light to propagate in the low-loss gaseous clagging that is encapsulated inside of a certain mechanical superstructure. Moreover, versatility, complexity, and cost of the fiber fabrication techniques are important factors for their ultimate acceptance by the industry and for practical deployment. While the gold standard for fabrication of the kilometer-long optical fibers is a drawing technique, THz range offers additional opportunities for fiber manufacturing as one needs only several tens of meter long THz fibers. In this paper we argue that 3D printing, and its variant - an infinite 3D printing, offer a promising fabrication method for multimeter long fibers with complex crossections.

We now briefly review some of the popular fabrication techniques of THz waveguides. The THz waveguides are generally classified into two types: metallic and dielectric. The metallic THz waveguides suffer from high ohmic losses that ultimately limit the link distance. Such waveguides are often used in integrated circuits and short-range interconnects [17, 18]. On the other hand, dielectric THz waveguides, particularly, the Microstructured Polymer Optical Fibers (MPOFs) are capable of guiding over tens of meters distances [19]. Normally, the MPOFs are fabricated using a fiber drawing technique that controllably softens a fiber preform, which is then pulled into a fiber. With the recent advance of the Rapid Prototyping techniques, such as Stereolithography (SLA) and Fused Deposition Modeling (FDM), the direct printing of complex MPOFs has been gaining popularity as such techniques eliminate the need for expensive fiber drawing infrastructure, as well as development and finetuning of complex preform fabrication and fiber drawing processes.

Particualry, the SLA technique uses layer-by layer selective photopolymerization, offers high resolution (~50 μm) and is capable of making structures of virtually unlimited 3D complexity. Due to high resolution of the SLA technique, fabricated THz structures can operate even at higher frequencies ~0.5–1 THz as the corresponding wavelengths (600–300 μm) are still much longer than the SLA resolution. At the same time, photopolymer resins used in this process are relatively lossy (>10 cm$^{-1}$) in the THz range, consummables are expensive, while build volumes are limited to ~10cm in every direction, thus allowing fabrication of only short waveguide sections. Due its high spatial resolution, the SLA technique was successful in demonstration of many advanced waveguides such as Kagome photonic crystal hollow-core waveguides [20], pentagram hollow-core anti-resonant waveguides [21], double pentagonal nested hollow-core waveguides [22], hollow photonic bandgap waveguides with hyperuniform disordered reflectors[23], two-wire plasmonic THz circuits [24], hollow-core Bragg waveguides with integrated fluidic channels [25], and many others.

An alternative to the SLA technique is an FDM technique that lays out layer-by-layer of a thermopolymer melt squeezed out from a hollow nozzle. Due to thermo-mechanical nature of the process, FDM resolution is limited by the nozzle opening that normally exceeds 200 μm. At the same time, FDM techniques can use a variety of polymer materials that in the THz range (0.1-0.3 THz) featuring medium-losses (~0.1-5 cm$^{-1}$) Poly(methyl methacrylate) (PMMA) [26], Polyethylene Terephthalate Glycol (PETG) [27], Acrylonitrile Butadiene Styrene (ABS) [28, 29], Polycarbonate (PC) [30], Polylactic Acid (PLA) [1], or low-losses (~0.01-0.5 cm$^{-1}$) such as Polystyrene (PS) [31], High Density Polyethylene (HDPE) [32], Cyclic Olefin Copolymer (also known as TOPAS)[33] and Polypropylene (PP) [15, 34]. Moreover, build volumes of FDM systems can be as big as 1m in every direction. Therefore, FDM technique is considered promising for the fabrication of THz waveguides and components for operation at lower THz frequencies ~0.1–0.3 THz, where operation wavelengths (~3–1 mm) are still much longer than the FDM resolution (~200 μm). Many MPOFs have been recently demonstrated using FDM technique. Among those, the solid-core THz waveguides are the easiest to fabricate, while their transmission losses are typically high and close to those of the fiber materials. However, by using low-loss polymers (ex. PP) and subwavelength size core, the modal transmission losses can be greatly reduced by pushing the modal field out of the lossy core and into the low-loss air cladding [15, 19, 35]. This enables the fabrication of high-performance THz waveguides and components for THz communications. Furthermore, subwavelength solid-cores can be encapsulated inside of the hollow support structure to reduce the effect of the environment and to improve the convenience of handling [36].

One of the key limitations of standard 3D printers for THz fiber development is a limited build volume. Recently, a novel additive manufacturing approach known as infinite 3D printing was developed that enables fabrication of 3D structures without any length limitations along a single direction. This method opens a possibility of fabricating THz fibers or even complete fiber devices of unlimited length and arbitrary complex 2D and 3D profiles. In this work, we aim at performing a comparative analysis of standard and infinite FDM printing to manufacture long and geometrically complex fibers for THz communication applications. Particularly, we fabricate a microstructured fiber that features a subwavelength core suspended in the air by three thin bridges in the middle of an encapsulation tube that allows convenient handling of such fibers without perturbing the modal fields. Moreover, the fiber is designed to operate at the point of zero dispersion at the carrier frequency of 128 GHz, and is made of a low-loss Polypropylene polymer. This fiber structure is chosen as a benchmark for the comparative analysis of FDM techniques as it features complex transverse geometry (thin bridges, large encapsulation tube), complex guidance mechanism (total internal reflection in the core and radiation leakage through the bridges), complex dispersion managed design, and challenging material (PP) for FDM technique. At the end we find that, indeed, the newly developed infinity FDM technique is well capable of fabricating such complex fibers, while significant investment into manufacturing process improvement and optimization are still necessary to make their performance compatible to standard FDM printed fibers.

The paper is organized as follows: Section 2 introduces the design of suspended-core microstructured fiber followed by the parameter optimization for 3D printing with PP polymer. Then, the fabrication process of the proposed fiber using standard and infinity 3D printers is detailed. Section 3 present the theoretical study of the fiber modal structure and their excitation efficiencies, modal optical properties including straight and bent fiber losses, group velocity dispersion, as well as projected information capacity of the fiber links. Finally, section 4 details experimental characterization of the fiber optical properties, modal imaging, as well as Bit Error Rate measurements of various fiber links as a function of the data bitrate

## 2. Fiber design and fabrication with standard and infinity 3D printers

The schematic of the proposed fiber is shown in Fig. 1. The blue regions represent the PP material, while the white regions are air. The fiber design was optimized to enable low-loss, near zero dispersion operation at the $v_c = 128\ GHz$ carrier frequency, which is the frequency of optimal performance of our THz communications setup detailed [15] (see Supplementary Materials Section S.1 for details on fiber optimization). Thus optimized fiber features a negative curvature solid core of inscribed circle diameter of ~ 1.61 mm suspended by three supporting bridges of $H_{br} = 0.4\ mm$ width. The cladding region is formed by three air holes with radii $R1_{cr} = \sim 3.7\ mm$, which are symmetrically distributed around the fiber center. The distance between the fiber center and the air hole center is $R2_{cr} = 4.5\ mm$. The outer fiber diameter is 9.0 mm, and the cladding thickness is $H_{cl} = 0.2\ mm$.

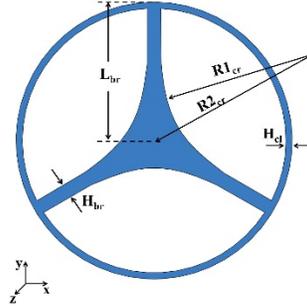

Fig. 1. Schematic of the cross-section of a suspended core fiber.

In what follows we use the Fused Deposition Modeling (FDM) to fabricate THz fibers using Polypropylene plastic that features one of the lowest losses in the THz spectral range and is translucent in the visible. The first task when 3D printing THz fibers is to optimize the printing process by minimizing the amount of trapped air in the fiber bulk regions, and to reduce surface roughness. After a comprehensive printing process optimization (see Supplementary Materials Section S.2 for details of printing process optimization), the following optimal PP printing parameters were found: infill flow rate of 110%, infill speed of 30 mm/s, layer height of 0.15 mm, first layer printing speed of 50 mm/s, inner shell printing speed of 70 mm/s, overlap of 5%, extrusion temperature of 240 °C and the built plate temperature of 95 °C.

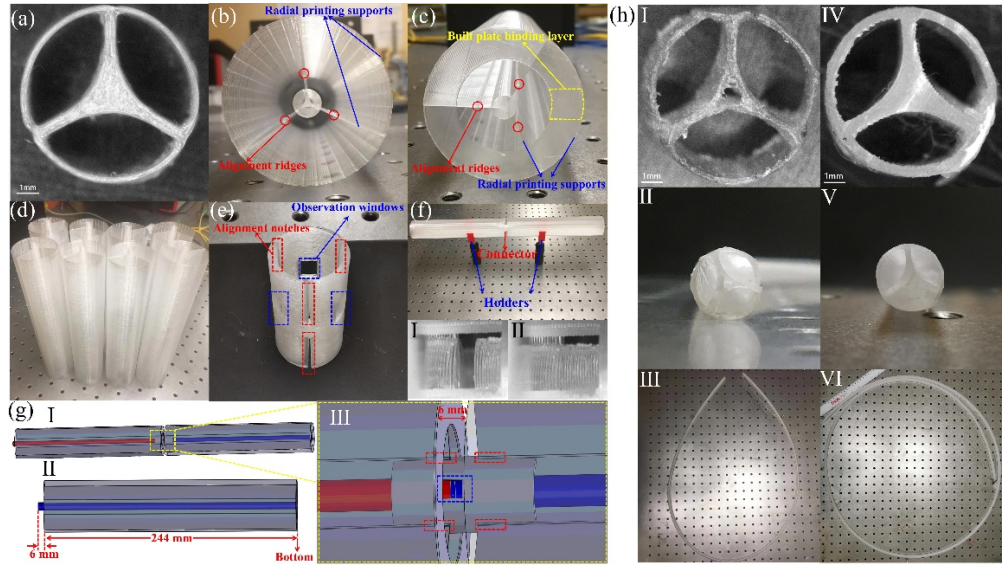

Fig. 2. (a) Cross-section of the FDM printed fiber with printing supports removed. (b) As printed fiber section (top view) with radial printing supports and inter-section alignment ridges. (c) As printed fiber section (bottom view) with radial printing supports, a built plate binding layer, and alignment ridges. (d) Eight of the 25 cm-long fiber sections to be combined into a 2 m-long fiber. (e) The connector and its 3D schematics. (f) Two fiber sections connected using one connector and two holders. Inserts: I (disconnected) and II (connected) - microscope view of the connection

between two fiber sections. (g) 3D Schematics of (I) the interconnection between two fiber sections and (II) one single fiber section. (III) An enlarged schematic of the connector region featuring alignment elements (red dotted regions) and observation windows (blue dotted region). (h) Two several meter long fibers printed using infinity FDM printer. (I) (IV) Microscopy images of the fiber cross-sections, (II)(V) side views and (III)(VI) top views of the continuous 1.4 m-long (defect core) and 2.5 m-long (solid core) fiber sections.

Fig. 2(a) presents the microscope image of the fiber cross-section printed using FDM technique and design parameters mentioned earlier. We refer to such fibers as "standard" in the rest of the paper. The 25 cm-long fiber section is printed with a printer in vertical direction aligned with the fiber length. To mechanically stabilize printing tall slender sections, a thin (0.2 mm) outer shell and three thin bridges (0.4 mm) are added on the outside of the fiber. The outer shell bridges are made 6 mm shorter than the fiber core (25 cm) in order for the cores of the two adjacent fibers to touch inside of the fiber connector shown in Figs. 2(b) and 2(g). While, the mechanical support structure can be easily cut from the fiber, we leave it in place to provide alignment and support during experimental measurements. An annular built plate binding layer at the bottom of the model is used to further stabilize printing [Fig.2(c)], while PP tape is applied to the built plate to promote better adhesion. Finally, thus printed 8 sections [Fig. 2(d)] are assembled into a 2 m-long waveguide using mechanical splicing with the help of separately printed connectors [Fig. 2(e)] that fit on the outside of the fiber cladding. To this end, each side of the fiber sections features three alignment ridges [Figs. 2(b) and 2(c)] that fit the appropriate notches in the connector, resulting in seamless transitions from one fiber section into the other with cores, bridges and cladding aligned up to ~50 microns positional precision with respect to each other. In Figs. 2(g) and 2(f) we show schematics of the two joined waveguide sections and a connecting region.

When using standard FDM printers, fiber length is limited by the printer linear dimensions (usually smaller than 0.5 m), while longer fiber sections must be assembled from shorter individually printed sections using connectors. Recently, a new class of FDM printers known as "infinity printers" has been developed. Such printers are designed to produce longer parts, or provide continuous manufacturing of smaller parts. They use continuous printing on a slowly moving belt (which serves as a build plate) using a 45° inclined extruder. In collaboration with BlackBelt 3D BV Inc., two PP fibers (1.4 m and 2.5 m) were fabricated using their flagship infinity printer with the fiber geometry presented earlier [Figs. 2(h) I and IV], and materials processing parameters similar to those used in a standard FDM printer. Images of the resultant fiber cross-sections, as well as fiber side and top views are shown in Fig. 2(h). The first 1.4 m-long fiber was printed directly on the belt without using any support and resulted in a small hole inside the fiber core; we refer to such a fiber as a "defect core" in the rest of the paper. The second 2.5 m-long fiber was printed by adding a thin rectangular binding layer between a fiber and a belt, thus improving the positional stabilization of the printing process; we refer to this fiber as "solid core" in the rest of the paper. For the two defect/solid core fibers their outer diameters are ~8.2/~8.0 mm, their inner cladding diameters are ~7.7/~7.0 mm, the smallest thicknesses of the three bridges for the two fibers are ~0.5/~0.6 mm, while their corresponding core sizes are ~1.5/~1.9 mm.

In what follows we aim at showing that both standard FDM-printed and properly connectorized fiber sections, as well as single strands of infinity FDM-printed fibers can be used to realize multi-meter THz fiber-assisted communication links. Both of these methodologies have their own relative advantages and disadvantages. At this point, it seems that while using infinity printing, the main advantage is in its ability of a single step fabrication of long continuous fiber links, at the same time the quality of its prints seems to be inferior to those of standard FDM printers with further thorough optimization of the infinity printing process for THz fiber manufacturing still in order.

## 3. Theoretical analysis of fibers fabricated using standard and infinity 3D printing

The following acronyms are used in the rest of the paper to simplify notation. Thus, fibers manufactured using infinity FDM printing are noted as either infinity defect core fiber (IDCF) or infinity solid core fiber (ISCF), while the ones fabricated using standard FDM printing are called standard suspended core fibers (SSCF). The dimensions of the three fabricated fibers are summarized in Table 1.

**Table 1. Fiber Dimensions**

| Fiber | Outer Diameter | Inner Diameter | Bridge Width | Core size |
|-------|----------------|----------------|--------------|-----------|
| SSCF  | 8.0 mm         | 7.6 mm         | 0.4 mm       | 1.6 mm    |
| IDCF  | 8.2 mm         | 7.7 mm         | 0.5 mm       | 1.5 mm    |
| ISCF  | 8.0 mm         | 7.7 mm         | 0.6 mm       | 1.9 mm    |

### 3.1 Modal structure of the straight fibers

The three experimentally realized fibers were numerically studied using finite element COMSOL Multiphysics software. The ideal cross-section shown in Fig. 1 was used to model optical properties of the SSCF. For the IDCF and ISCF, the two-dimensional models of the fiber cross-section were built using high-resolution microscope images of the corresponding fiber cross-sections imported into COMSOL. For all the fibers, we assume that the cladding is air. Effective refractive index $n_{PP}$ and absorption coefficient by power $\alpha_{PP}$ of the Polypropylene plastic were taken from our prior study [15], and in the frequency range of $\nu$ $(0.1 - 0.15\ THz)$ they can be fitted as $n_{PP} = 1.485\ and\ \alpha_{PP}\ [dB/m] = 236.31\ \nu^2 - 37.75\ \nu + 3.32$, where frequency $\nu$ is in $[THz]$. In our simulations, we focus on the fiber key optical parameters such as modal loss, excitation efficiency, bending loss, and group velocity dispersion that directly impact link transmission length and link bitrate.

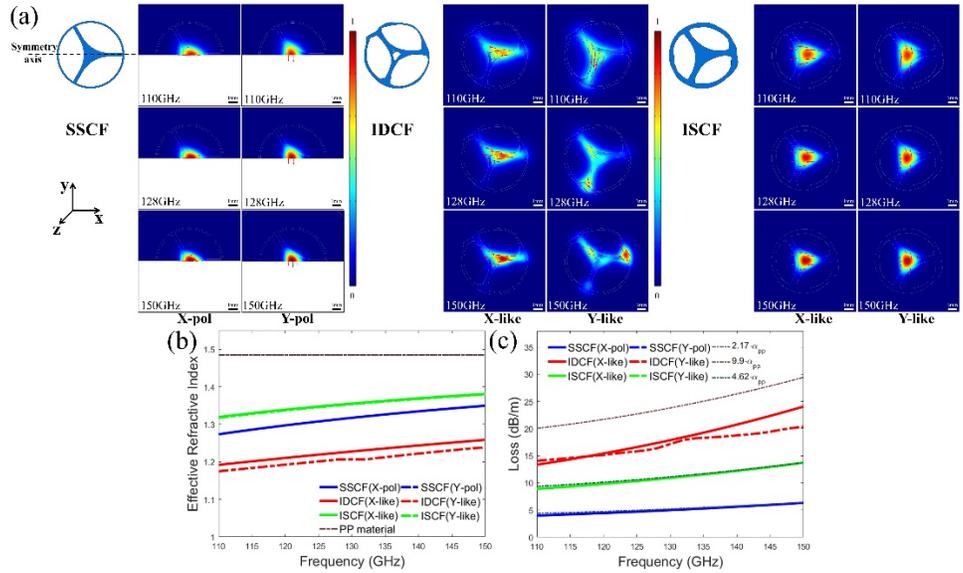

Fig. 3. (a) Normalized electric field distribution and electric field direction (red arrows) of the two lowest order modes for the three fibers as a function of frequency. (b) Theoretical effective refractive indexes and (c) transmission losses (by power) for the two lowest order modes of three fibers as a function of frequency. Dotted lines show corresponding effective fiber material losses.

First, we note that the ideal fiber shown in Fig. 1 supports truly doubly degenerate fundamental modes, while experimental fibers shown in Figs 2(h) and 3 support the nearly double degenerate fundamental modes. Generally, a certain linear combination of the degenerate or near degenerate modes will be excited at the fiber input depending on the excitation conditions. For practical applications, however, it is beneficial to optimize coupling conditions in order to preferentially excie a single mode in order to mitigate the negative effects of the inter-modal interference and inter-modal dispersion that can affect fiber information transmission characteristics (see Supplementary Materials Section S.3 for detailed explanation of the near degeneracy in experimental fibers). Before we address optimization of the modal excitation efficiency, we first study the modal structure for the three abovementioned fibers in the frequency range of 110 – 150 GHz using geometries shown in Fig. 3(a). The normalized electric field distributions and electric field directions (red arrows) for the two lowest order modes for the three fibers are presented in Fig. 3(a). Only in the case of SSCF, the X and Y polarizations can be unambiguously defined using reflection symmetries. Particularly, in the left panel of Fig. 3(a) we show field distributions of the properly symmetrized doubly degenerate fiber modes for SSCF, where X-polarized mode is calculated using half computational-cell and the Perfect Magnetic Conductor (PMC) boundary, while the Perfect Electric Conductor (PEC) boundary is used for calculating the Y-polarized mode. In the case of IDCF and ISCF, the two lowest order non-degenerate modes are calculated using a full computational-cell and their field distributions are presented in the middle and right panels of Fig. 3(a). In what follows, we classify the lowest order modes of IDCF and ISCF as X-like and Y-like by analogy with the modes of an SSCF after inspection of their corresponding electric field distributions.

The effective refractive indexes of the two lowest order modes for the three fibers are presented in Fig. 3(b) and all show monotonic increase at higher frequencies. This is due to higher modal confinement in the waveguide core at higher frequencies as seen from the modal field distributions shown in Fig. 3(a). Moreover, at any given frequency the ISCF has the highest effective refractive index as it features the largest core size. In contrast, the IDCF fiber has the smallest effective refractive index due to much weaker modal confinement caused by the air hole in the core center.

In Fig. 3(c) we present the calculated modal transmission losses, as well as adjusted bulk losses of the fiber material as a function of frequency. When calculating fiber losses we assume a frequency dependent fiber material loss proportional to that of a bulk Polypropylene (from [15]) up to a multiplicative factor, which we choose to best reproduce the experimentally measured fiber transmission losses (see Section 4.1). Thus, for SSCF, for the fiber core material loss (by power) we used $\sim 2.17 \cdot \alpha_{PP}$, for the IDCF fiber we used fiber material loss of $\sim 9.9 \cdot \alpha_{PP}$, and for the ISCF we used $\sim 4.62 \cdot \alpha_{PP}$. The multiplicative factors were found using the least-squares method to minimize the fitting error between the experimental and numerical loss data. The fact that the two infinity-printed fibers have much higher fiber material losses than the bulk PP material is attributed to high scattering loss caused by the roughness of various fiber surfaces due to printing process. Also, we note that standard FDM printing is highly optimized and, thus, results in the smallest scattering loss among all the fibers, while further work is in order to optimize infinite-FDM process to further reduce scattering loss due to manufacturing process.

For SSCF and ISCF, the two lowest order modes feature losses that are close to the effective fiber material losses since both modes are well confined inside the fiber solid core. In contrast, for IDCF, both lowest order modes are strongly present in the air holes inside and surrounding the fiber core, thus the modal losses are much smaller than the effective fiber material losses. Specifically, losses (by power) of the two almost degenerate lowest order modes for the SSCF in the range of 110 – 150 GHz are found to be 3.99 – 6.34 dB/m, with the corresponding value at the 128 GHz carrier frequency being 4.88 dB/m. For IDCF and ISCF, losses of the X-like modes in the range of 110 – 150 GHz are 13.36 – 24.06 dB/m for IDCF, and 8.93 –13.72 dB/m for ISCF, with 17.35 dB/m (IDCF) and 10.7 dB/m (ISCF) losses at 128 GHz. Similarly, losses of the Y-like modes for the two fibers in the 110 – 150 GHz frequency range are 14.08 – 20.34 dB/m (IDCF) and 8.9 – 13.71 dB/m (ISCF), with the corresponding values at 128 GHz of 16.38 dB/m (IDCF) and 10.68 dB/m (ISCF). Furthermore, we also studied bending losses of the X-polarized or X-like modes of three bent fibers for two orthogonal fiber bending directions at 128 GHz, which shows that all the three fibers can readily tolerate tight bends with radius as small as 3 cm, resulting only in small loss increase of <0.01 dB per 90° bend (see Supplementary Materials Section S.4 for numerical analysis of fiber bending losses).

*3.2 Excitation efficiency of the fiber modes*

Here we study excitation efficiencies of the fiber modes using WR-6 waveguide flange as a source. The flange supports a single linearly polarized mode (along X direction in the experiments) with transverse electric field directed along the shorter side of a rectangular metallic waveguide. The complex modal excitation coefficient can be estimated using a well know expression[37]:

$$C_m = \frac{\iint (E_{mode}^* \times H_{wg} + E_{wg} \times H_{mode}^*) dx dy}{\sqrt{\iint 2Re(E_{wg}^* \times H_{wg}) dx dy} \sqrt{\iint 2Re(E_{mode}^* \times H_{mode}) dx dy}} \quad (1)$$

where $E_{mode}$ and $H_{mode}$ are the transverse electric and magnetic fields of a given fiber mode, while $E_{wg}$ and $H_{wg}$ are the transverse electric and magnetic fields of the WR-6 waveguide flange. Then, relative power excited in the waveguide mode is given by $|C_m|^2$ which we refer to as excitation efficiency in the rest of the paper.

We now perform optimization study aimed at selective excitation of a single mode in each of the three waveguides. Experimentally, the WR-6 flange as shown in Fig. 4(c) with a fixed transverse electric field direction along the X-axis was used for fiber excitation. In our experimental studies we favor X-like modes to the Y-like modes, especially for the IDCF.

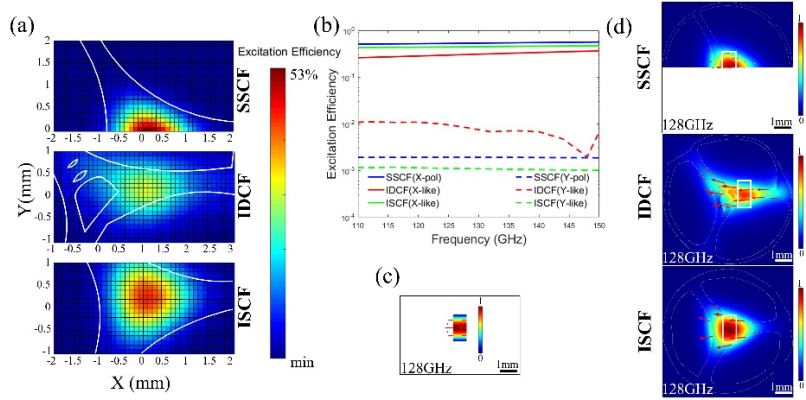

Fig. 4. (a) Excitation efficiencies for the X-polarized and X-like modes for SSCF (top), IDCF (middle) and ISCF (bottom) as a function of the center position of the WR-6 waveguide flange at 128 GHz. WR-6 mode is X-polarized. Shown excitation efficiencies are optimized with respect to the fiber rotations. The white solid lines are fiber boundaries. (b) Excitation efficiencies of the two lowest order modes of the three fibers versus frequency for optimized coupling arrangement. Coupling is optimized for X-polarized or X-like modes at the carrier frequency of 128 GHz. Both fiber rotation and fiber position is optimized relative to a fixed WR-6 waveguide flange; Normalized electric field distributions and electric field directions (red arrows) for (c) the X-polarized mode of the WR-6 waveguide flange at 128 GHz and (d) the X-polarized or X-like modes of the three fibers at 128 GHz. White rectangles show optimized positions of the WR-6 waveguide.

To maximize the excitation efficiencies of the X-like modes for the IDCF and ISCF, one has to optimize both the fiber inclination and the relative position between the fiber and WR-6 waveguide. For the X-polarized SSCF mode, one can forgo the fiber rotation optimization step and simply choose polarization of the WR-6 waveguide mode along the fiber reflection symmetry axis (see Supplementary Materials Section S.3 for details of the selective excitation optimization procedure). The map of excitation efficiencies for the X-polarized mode and X-like modes for the three fibers as a function of the center position of the WR-6 waveguide flange operating at 128 GHz are shown in Fig. 4(a). These excitation efficiencies are also optimized with respect to the fiber rotations, assuming that polarization of the WR-6 waveguide mode [shown in Fig 4 (c)] is fixed along the X direction. The maximal excitation efficiencies of the X-polarized and X-like modes at the carrier frequency of 128 GHz for the SSCF mode reaches ~53% , while for the IDCF and ISCF they reach ~31% and ~45% respectively. The corresponding optimal fiber orientations and positions with respect to the WR-6 waveguide flange (indicated as white rectangles) are shown in Fig.4 (d) along with electric field distributions of the X-polarized and X-like modes. Thus optimized coupling efficiencies for the excitation of X-polarized and X-like modes over the frequency range of 110 – 150 GHz are shown in Fig. 4 (b), with excitation efficiencies for the SSCF ranging between ~ 51% – 57%, and those for the IDCF and ISCF ranging between ~ 26% – 37% and 44% – 47%, respectively. Finally, in order to quantify efficiency of a single mode excitation using the optimized procedure described earlier, in Fig. 4(b) we also present excitation efficiencies for the lowest-order Y-polarized and Y-like modes calculated for the optimal excitation conditions of the X-polarized and X-like modes, and note that they are all below 1%. This confirms that optimized excitation procedure developed in our work guarantees an effectively single mode excitation with 14 – 27 dB suppression (by power) of other lowest order modes in all the three fibers.

*3.3 Modal Group Velocity Dispersion and maximum bit rate*

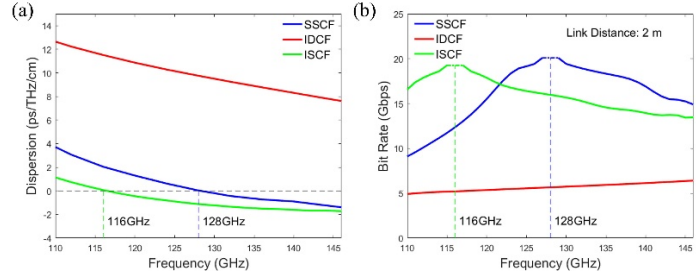

Fig. 5. (a) Second-order dispersion of the X-polarized and X-like modes versus frequency for the three fibers. (b) Maximum bit rate supported by the three fibers for a link length of 2 m.

Another important factor affecting signal quality transmitted through optical fibers is dispersion. While the intensity of a received signal should be significantly above the receiver detection level, it is also essential to minimize signal distortion due to dispersion. The dispersions of X-polarized or X-like modes for the three fibers as a function of frequency are shown in Fig. 5(a). The dispersion of SSCF at the carrier frequency of 128 GHz is near zero by design while it is in the range of –2 to 4 ps/THz/cm in the whole 110 – 150 GHz frequency range. For ISCF, the dispersion curve is similar to that of a SSCF with the zero-dispersion frequency somewhat shifted to 116 GHz due to deviation of the fiber geometry from the optimal one during manufacturing. Finally, dispersion of the IDCF is large and positive ~10 ps/THz/cm in the whole frequency range due to a hole defect in the fiber core. Assuming an infinite Signal to Noise Ratio (zero noise), the maximal bit rate supported by a single mode fiber at a given carrier frequency for a simple ON-OFF keying modulation can be estimated as [38]:

$$BR_{max} = 1/\left(4\sqrt{|\beta''|L}\right) \quad (3)$$

where $\beta''$ is the second order derivative of the modal propagation constant, and $L$ is the fiber length. At the frequency of zero dispersion the maximum error-free bit rate can be estimated using the third order modal dispersion [38]:

$$BR_{ZD} = 0.324/\sqrt[3]{|\beta'''|L} \quad (4)$$

where $\beta'''$ is the third order derivative of the modal propagation constant. In Fig. 5(b) we plot the maximal estimated bitrate $BR_{max}$ as a function of the carrier frequency, while capping its maximal value by $BR_{ZD}$, while assuming a $L = 2\ m$ fiber length. From Fig. 5 (b) we see that in the vicinity of the corresponding zero dispersion frequencies, the 2 m-long SSCF and ISCF can support error-free transmission bit rates over 20 Gbps (at 128 GHz) and 19 Gbps (at 116 GHz). Meanwhile, at 128 GHz the 2 m-long IDCF and ISCF can support the error-free transmission rates of up to ~5 Gbps and ~16 Gbps, which are equally attractive for various short-range fiber-assisted communication applications.

## 4. Experimental characterization of the THz fiber communication links

The experimental characterizations were carried out using an in-house photonics-based THz communication system detailed in [15]. The same system can be used both in the Continuous Wave (CW) THz spectroscopy and THz communication mode by simply disabling or activating the communication unit (see Supplementary Materials Sections S.5 and S.6 for details of the two systems).

### 4.1 Transmission loss measurements

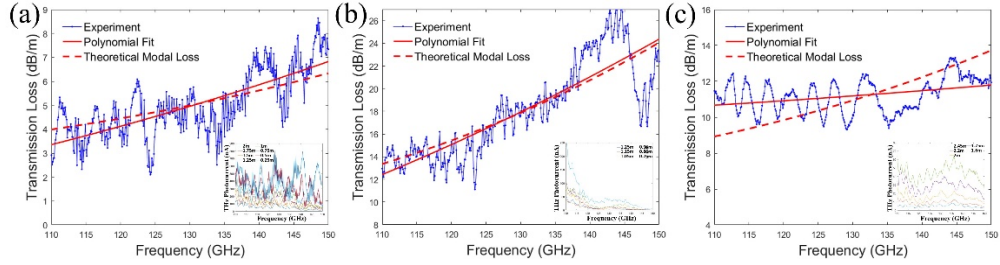

Fig. 6. Experimental transmission losses (by power) of the X-polarized and X-like modes for (a) SSCF, (b) IDCF and (c) ISCF. Insert: experimental transmission spectra. Firstly, the transmission losses of the 3D printed fibers presented above were measured using the standard cut-back method. The measurements were carried out with the CW THz spectroscopy system (see Supplementary Materials Section S.5 for details). A total of 8, 6 and 5 transmission spectra in the range of 110 – 150 GHz were obtained respectively for the SSCF, IDCF and ISCF fibers [see Figs. 6 (a-c) Inset]. The maximal length of each fiber link was mostly limited by the fiber losses at all frequencies (see Section 4.2 for details). At a given frequency $\nu$, fiber loss $\alpha(\nu)$ is estimated by minimizing the least squares deviation of the experimentally measured transmitted intensities $I$ (measured as photocurrents) as a function of the fiber length from the theoretically expected one (5). In this fitting procedure, frequency dependence of the fiber loss is assumed to be second order polynomial:

$$I = I_0 \cdot exp(-\alpha(\nu)L)$$
$$\alpha(\nu) = a_2\nu^2 + a_1\nu + a_0 \quad (5)$$

Thus found fiber loss is shown in Figs. 6(a-c) as solid red curves. Similarly, the transmission losses of the X-polarized (SSCF) and X-like (IDCF and ISCF) modes are numerically calculated using COMSOL mode solver and are shown as dashed red curves on the same figures, and a good agreement with the measurements is observed. At the carrier frequency of 128 GHz, the measured fiber transmission losses (by power) are found to be 4.79 dB/m, 17.34 dB/m and 11.13 dB/m for the SSCF, IDCF and ISCF correspondently.

### 4.2 Bit error rate measurements

Next, the bit error rate (BER) measurement was carried out to evaluate the communication performance of the 3D printed fibers using the photonics-based THz communication system. The communication unit was enabled and the 3D printed fibers were coupled to the transmitter and receiver antenna in a similar arrangement to the modal loss measurement (see Supplementary Materials Section S.6).

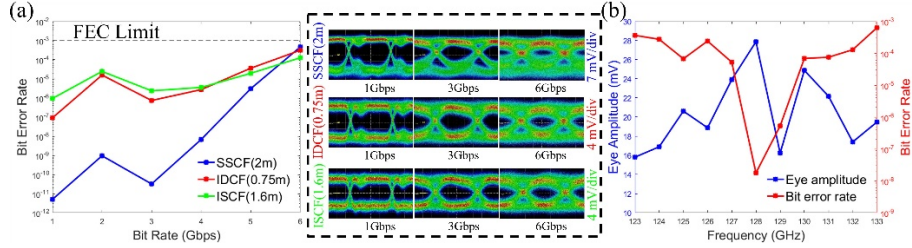

Fig. 7. (a) Measured Bit Error Rate versus Bit Rate for SSCF (2 m), IDCF (0.75 m) and ISCF (1.6 m) at the carrier frequency of 128 GHz. Insert: The recorded eye pattern for different bit rates. (b) Measured eye amplitude and bit error rate for SSCF (2m) at the bit rate of 4 Gbps as a function of frequency.

The SSCF was assembled from 3D printed sections of 25 cm each to the total length of 2 m, while longer fiber links can be readily achieved by connecting more sections. The maximal fiber lengths of IDCF and ISCF for conducting the BER measurement were limited to 0.75 m (IDCF) and 1.6 m (ISCF). The fiber lengths were chosen to result in similar total fiber link transmission losses of ~15 – 16 dB (which includes both coupling and fiber losses) as measured by the eye amplitudes of the oscilloscope. As the noise level of our THz communication system is -34 dBm(~2.5 mV) and the signal strength is -6.6 dBm, then, after fiber transmission we are still operating ~12 dB in power above the noise level. The BER and corresponding eye patterns at 128 GHz for the fibers at the bit rates of 1 – 6 Gbps are presented in Fig. 7 (a).

From Fig.7(a), we see that, the lowest bit error rates (BER < $10^{-8}$) were observed for the SSCF, particularly for the bit rates below 4 Gbps, due to fiber operation near frequency of zero modal dispersion. Although the fiber length is the longest among the three fibers (2 m), the eye amplitude measured for the received signal is larger than those for the IDCF and ISCF because of the SSCF lower transmission losses. We also note that, in the case of SSCF fiber, the bit error increases rapidly for higher bit rates. This happens because for higher bit rates, the channel bandwidth increases, and fiber can no longer be considered as operating at the frequency of zero dispersion. In optical communications one frequently uses dispersion flattened fiber designs to increase the bandwidth of low dispersion operation and a similar tactic can be used for advanced design of THz fibers. It is worth mentioning that, an error-free transmission (BER < $10^{-12}$) was achieved at the bit rate up to 5.2 Gbps with SSCF of 1.5 m-length. In comparison, for the IDCF and ISCF, the BER is much higher than for SSCF as they operate away from their respective zero dispersion frequencies. At the same time, for all three fibers, signal transmission with BERs below the forward error correction (FEC) limit ($10^{-3}$) is supported even at bit rates as high as 6 Gbps.

To demonstrate that we operate in the vicinity of a zero dispersion frequency of the SSCF fiber, we have conducted the BER measurement and recorded the eye amplitudes for the 2m-long SSCF fiber at the bit rate of 4 Gbps by varying the carrier frequency from 123 GHz to 133 GHz as shown in Fig. 7(b). From this figure we observe that by varying the carrier frequency, the eye amplitude (defined as a min to max value of a signal) varies from ~16 mV to ~28 mV, while achieving its maximal value at 128 GHz. While the eye amplitude drops suddenly at 129 GHz (which is solely due to response of our detector as established earlier [7]), nevertheless, the bit error rate (~$10^{-8}$) shows a clear minimum in the whole 127 - 130 GHz range, which we attribute to operation near zero dispersion frequency of 128 GHz as predicted theoretically in Fig. 5.

In conclusion, we project that, by resorting to dispersion flattened designs in the vicinity of zero dispersion frequency, and by optimizing the 3D printing quality, transmission rates of ~10-

20 Gbps in the 3D printed ~10 m-long THz fiber links are possible even when using low-power (~0.1mW) THz optical sources.

*4.3 Mode field imaging*

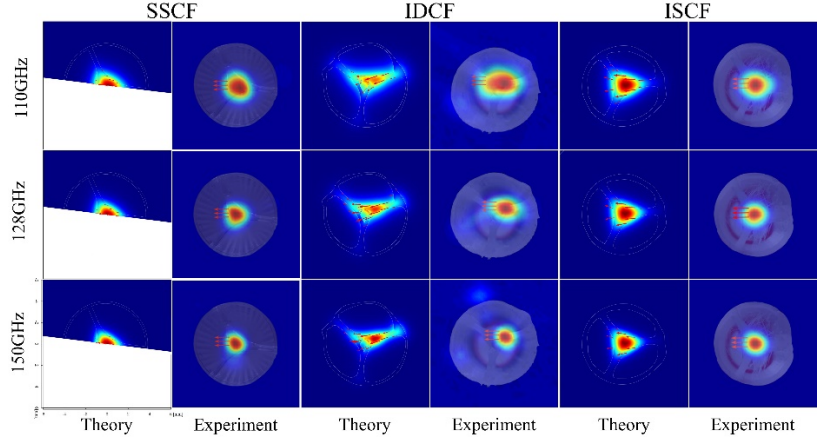

Fig. 8. Comparison of the theoretically and experimentally measured electric field distributions of the X-polarized and X-like modes of the SSCF (0.25 m), IDCF (0.75 m) and ISCF (1.6 m). Electric field directions are shown as red arrows. In experimental measurements, electric field direction is defined by the orientation of the WR-6 waveguide flange, which is horizontal.

To verify the single-mode guidance of the fabricated 3D printed fibers, a near-field THz modal imaging has been carried out at the output end of the fiber using the CW THz spectroscopy system (see Supplementary Materials Sections S.5 and S.7). The experimental and theoretical modal field distributions of the 3D printed fibers at three different frequencies, along with the electric field directions (red arrows), are presented in Fig. 8. The red arrows shown in the theoretical result are the electric field directions of the X-polarized or X-like modes guided in the three fibers, while arrows in the experimental figures show the electric field direction in the fundamental mode of the WR-6 waveguide flange used for waveguide excitation at 128 GHz. It should be noted again that the polarization directions of the transmitter and receiver antennas were always fixed to horizontal direction, while theoretical field distributions in Fig. 8 are rotated to match the fiber orientation of the experimental setup. Overall, there is a good correspondence between the theoretical and experimental modal images, while minor differences come from the fact that our THz imaging setup only measures a single horizontal field component of the electric field while averaging it over ~1mm aperture. Thus, one clearly observes stronger modal confinement in the core at higher frequencies for all three fibers, as well as field displacement into the bridge region due to hole defect in the IDCF core.

## 5. Conclusion

In this work, we explored an infinity 3D printing technique to fabricate continuous several-meter-long low-loss near-zero dispersion suspended-core polypropylene fibers for application in terahertz communications. The suspended-core geometry was chosen to shield the mode from external influence and simplify handling of such fibers in practical applications. While Polypropylene polymer was chosen as one of the lowest loss plastics in the THz regime, in fact, it is rarely used with FDM technique due to heavy warping of the material during printing. Therefore, a particular attention was paid to process parameter optimization for printing with low-loss polypropylene plastic, as well as in-depth comparison between three fibers printed using standard and infinity 3D printers.

Experimentally, the transmission losses (by power) of the three fibers produced using standard and infinity FDM techniques were measured to be 4.79 dB/m, 17.34 dB/m and 11.13 dB/m for the SSCF, IDCF and ISCF respectively at the carrier frequency of 128 GHz. Subsequently, the BER measurements were carried out by varying the bit rate between 1 – 6 Gbps for the three fibers. Signal transmission with BER far below the FEC limit was observed for the 2 m-long SSCF, 0.75 m-long IDCF and 1.6 m-long ISCF, respectively. Additionally, the BER measurements were conducted by varying the carrier frequency between 123 – 133 GHz for the 2 m-long SSCF at the bit rate of 4 Gbps, and a clear minimum in the BER (~$10^{-8}$) was observed at 128 GHz, which is a zero dispersion frequency chosen for the SSCF design. Moreover, an error-free transmission was achieved for the bit rate up to 5.2 Gbps using SSCF with a length of 1.5 m which can already be interesting for practical applications.

Finally, the near-field imaging of the fiber fields was performed by raster scanning of the fiber output facets with a sub-wavelength aperture. It showed strong modal confinement in the suspended-core region well inside of the outer protective shell.

We believe that our work has demonstrated that infinity 3D printing holds strong potential for development of THz fibers and fiber components via single step fabrication of unlimited in length fibers featuring complex geometrical cross-sections, while using low-loss plastics.

## Funding

## Disclosures

The authors declare that there are no conflicts of interest related to this article.

# Microstructured fiber links for THz communications and their fabrication using infinity printing: supplemental materials

**GUOFU XU, KATHIRVEL NALLAPPAN, YANG CAO, AND MAKSIM SKOROBOGATIY***

*Department of Engineering Physics, École Polytechnique de Montréal, Montreal, Québec, H3T 1J4, Canada*
*\*maksim.skorobogatiy@polymtl.ca*

## 1. Theoretical Fiber Structure Optimization

The optimization of the fiber structure was conducted using finite element COMSOL Multiphysics software. Half computational cell with Perfect Magnetic Conductor (PMC) boundary conditions set along the symmetry reflection plane was used to study the lowest order mode featuring electric field directed preferentially along the plane. The effective refractive index and absorption coefficient (by power) of the Polypropylene polymer were fixed at 1.49 and 0.2 cm$^{-1}$ (~ 8.69 dB/m). The fiber outer cladding was first assumed to be infinite plastic in order to quantify core to cladding radiation (leakage) loss due to finite bridge thickness $H_{br}$ and length $L_{br}$. The radius $R1_{cr}$ of the three air holes was defined as $R1_{cr} = R2_{cr} \cdot \sin(\pi/3) - H_{br}/2$. The distance $R2_{cr}$ was varied between 2.5 – 6.0 mm with a step value of 0.1 mm, in order to realise fibers of core size $2 \cdot (R1_{cr} - R2_{cr})$ ranging from ~0.87 mm to ~2 mm, and tune the spectral position of the zero dispersion frequency. The bridge length $L_{br}$ that controls coupling of the core and cladding regions was varied between 2.0 - 4.5 mm with a step value of 0.5 mm. The values for $H_{br}$ were set as 0.2 mm or 0.4 mm due to practical limitations when fabricating using 3D printing and to provide sufficient mechanical strength for the suspended core with minimal deformations. Therefore, by varying $R2_{cr}$ while fixing the values for $H_{br}$ and $L_{br}$, the fiber design featuring near zero dispersion at 128 GHz can be obtained. By comparing these dispersion optimized fibers featuring different parameter combinations of $H_{br}$ and $L_{br}$, we then found that the total fiber losses are comparable for the fibers with $L_{br}$ = 4.5 mm - 4 mm, while increasing rapidly ( by a factor of 2-3) when using shorter bridge lengths $L_{br}$=3.5 mm - 2mm. Therefore, the optimal $L_{br}$ was set to 4 mm in order to minimize the fiber core to cladding leakage loss, while still keeping the outer fiber size as small as possible to maintain the fiber flexibility. Furthermore, for the fibers featuring $L_{br}$ = 4 mm and $H_{br}$ = 0.4 mm or 0.2 mm, we found that while the one with $H_{br}$=0.2 mm has marginally smaller loss, however consistent printing with such a thin layer of internal fiber microstructure was problematic. Thus, the optimal $H_{br}$ was identified as 0.4 mm, with the corresponding values of $R2_{cr} = 4.5\ mm$ and $R1_{cr} = \sim 3.7$. Finally, during manufacturing, the cladding thickness $H_{cl}$ was chosen as 0.2 mm as it has enough mechanical strength to work as a robust mechanical shell for the fiber, while being practical to fabricate using 3D printing.

## 2. Optimization of the print quality using standard 3D printer

FDM technique unavoidably results in surface roughness with sizes comparable to the deposited layer thickness. Additionally, when printing the fiber bulk regions with 100% filling by volume, accidental air trapping is possible. Both surface roughness and air trapping in the bulk lead to additional scattering loss. At the same time, when operating in the low THz frequency range (most suitable for communication applications) the wavelength size is typically ~1 mm, while the layer thickness of an FDM printer, as well as its transverse resolution are in the ~0.1 mm range; thus surface and bulk roughness are deeply subwavelength, and scattering from such defects should obey the Rayleigh law.

Therefore, as an indicator to the print quality we use sample transparency in the visible. To achieve the best transparency (highest material uniformity) at 100% filling factor, dozens of cylindrical pellets (15 mm diameter and 4 mm height) were printed with somewhat different combinations of the printing parameters. In our studies we used "Raise3D Pro2" FDM printer that features high positional precision (~1 um in the XY plane and ~10um along Z axis), as well as high resolution in both horizontal (nozzle size of 0.2 mm) and vertical (minimal layer thickness of 0.1 mm) directions. The 1.75 mm natural transparent PP filament was employed for printing since it has high transparency and low absorption loss in the THz regime [1]. A standard "Rectilinear" infill pattern is used in the optimization process as it usually results in mechanically isotropic prints in the build plane. Optimal extrusion temperature for the Polypropylene is reported in the 190 – 250 °C range [2], and in our experiments is chosen to be 240 °C according to the filament manufacturer recommendations

Next, we perform a multiparameter optimization of the printing process for the Polypropylene material.

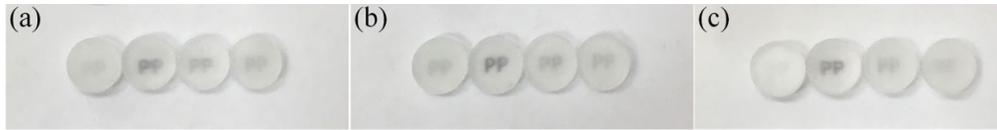

Fig. S1. (a) Pellets printed with different infill flow rates of 120%, 110%, 100%, 90% (decreasing from left to right). (b) Pellets printed with different infill speeds of 10 mm/s, 30 mm/s, 120 mm/s and 180 mm/s (increasing from left to right). (c) Pellets printed with different layer heights of 0.13 mm, 0.15 mm, 0.17 mm and 0.19 mm (increasing from left to right).

The Pellets printed with different infill flow rates of 90%, 100%, 110% and 120% at a constant infill speed of 30 mm/s and a fixed layer height of 0.15 mm are shown in Fig. S1(a). The optimal transparency is achieved at somewhat elevated infill flow rates of ~110%, as air gaps between the adjacent extruded lines tend to be minimized as extra amount material is squeezed out. However, when using higher flow rates, significant material overflow occurs leading to increase in surface roughness. The pellets printed with different infill speeds of 10 mm/s, 30 mm/s, 120 mm/s and 180 mm/s at a fixed flow rate of 110% and a fixed layer height of 0.15 mm are shown in Fig. S1(b). The best transparency is obtained at the relatively slow infill speeds of ~30 mm/s which tend to result in wider extruded lines and denser prints. However, if the infill speed is set too low, over-extrusion will occur ultimately increasing surface roughness and inhomogeneity of a print. Conversely, excessively high infill speeds will increase gaps between the adjacent lines, resulting in higher content of trapped air and poor transparency. Finally, in Fig. S1(c), we show pellets printed using layer heights of 0.13 mm, 0.15 mm, 0.17 mm and 0.19 mm printed at a fixed infill rate of 110 % and infill speed of 30 mm/s. From this figure we see that the layer height also makes significant impact on the transparency of a print, with an optimal layer thickness being 0.15 mm. While thinner printing layers generally provide higher printing quality and mechanically stronger prints, however, at very small layer thicknesses, fluctuation in the flow rate of a molten plastic can lead to significant variations in the layer thickness and, consequently, degradation of the print quality. In addition, several other parameters can influence the printing quality of certain structures, and, therefore, also need to be optimized. Thus, the combination of printing speed and infill overlap (with the walls) has a great impact on the size and quality of the printed slender parts, which is crucial for printing thin bridges in the fiber cross-section. Thinner shells and structures can be obtained when using higher printing speed and smaller infill overlap parameter. In addition, the heavy warping that often happens in the first few layers when printing with a PP filament can be addressed by using the appropriate first layer printing speed and built plate temperature.

## 3. Excitation of a fiber featuring doubly degenerate modes

In the case of a fiber supporting a doubly degenerate fundamental mode, or an almost degenerate pair of modes, the electromagnetic fields at distance $z$ along the fiber length can be expressed as:

$$F(x,y,z) = C_1 \cdot F_{1(x,y)} \cdot \exp(i\beta_1 z) + C_2 \cdot F_{2(x,y)} \cdot \exp(i\beta_2 z) \qquad (S1)$$

where $F_{1(x,y)}$ and $F_{2(x,y)}$ are the transverse electromagnetic fields of the two modes, $C_1$ and $C_2$ are the excitation coefficients of the two modes at the fiber input ($z = 0$), $\beta_1$ and $\beta_2$ are the propagation constant of the two modes. So, if the two modes are degenerate ($\beta_1 = \beta_2 = \beta$), the electromagnetic field propagating along the fiber will be:

$$F(x,y,z) = [C_1 \cdot F_{1(x,y)} + C_2 \cdot F_{2(x,y)}] \cdot \exp(i\beta z) \qquad (S2)$$

Thus, a linear combination of the degenerate modes will be excited at the fiber input depending on the external excitation source. The total power carried by the modes will be the output power of excitation source multiplied by $|C_1|^2 + |C_2|^2$. The same argument holds if the fiber supports almost degenerate modes ($\beta_2 - \beta_1 = \Delta\beta \ll |\beta_1|$) which is the case of SSCF fiber. Then, the electromagnetic field propagating along the fiber will be:

$$F(x,y,z) = [C_1 \cdot F_{1(x,y)} + C_2 \cdot F_{2(x,y)} \cdot \exp(i\Delta\beta z)] \cdot \exp(i\beta_1 z) \qquad (S3)$$

In this case, the two modes can be considered as degenerate if the propagation distance $z$ is smaller than $L_{\Delta\beta} = \pi/\Delta\beta$, with the total excited power $\sim |C_1|^2 + |C_2|^2$ and transverse field distribution $\sim [C_1 \cdot F_{1(x,y)} + C_2 \cdot F_{2(x,y)}]$. In the case of experimental fibers IDCF and ISCF, the $L_{\Delta\beta}$ is calculated to be $\sim 0.07$ m and $\sim 0.78$ m which are shorter than the fiber lengths of 1.4 m and 2.5 m used in the measurements. This means that for practical applications it is preferable to excite only one of the two modes to mitigate the negative effects of the inter-modal beating, inter-modal dispersion, etc. that can affect fiber information transmission characteristics. To excite our fibers, we use a fixed WR-6 waveguide flange that supports a single fundamental mode. Therefore, before conducting any communication measurements, one has to find the optimal fiber orientations that result in the most efficient excitation of a single preferred fiber mode.

In the case of a SSCF featuring a symmetry reflection plane (ZOX), its modes can be characterized as being Y or X polarized. In this case, excitation efficiency of a given mode is first optimized by aligning its principal polarization direction with that of a WR-6 mode. Then, by scanning the waveguide profile with a WR-6 flange along the symmetry axis, one finds the relative position of the two waveguides that maximizes coupling efficiency. The same approach largely holds when waveguide profile is almost symmetric, like it is the case of ISCF fiber shown in Fig. 3(a). Note that for a long-enough fiber, inter-modal scattering due to imperfections along the fiber length can lead to a significant power transfer between the modes of two distinct polarizations. This effect is especially pronounced in fibers that feature a doubly degenerate or a nearly degenerate fundamental mode like it is the case for SSCF and ISCF [see Fig. 3(b)]. Once a particular polarization is excited, the fiber will maintain its polarization state only over a certain maximal propagation distance that depends strongly of the imperfection strength and is difficult to predict analytically.

In the general case of a non-symmetric profile [IDCF fiber or ISCF fiber shown in Fig. 3(a)], the two lowest order fiber modes can no longer be described as X or Y polarized. Thus, both the fiber orientation with respect to the polarization direction of the WR-6 modal field, as well as relative position of the two waveguides have to be optimized simultaneously to maximize coupling efficiency into a single fiber mode. Additionally, X-like modes show stronger than Y-like modes confinement in the lossy core, and as a consequence they feature somewhat higher absorption losses, but at the same time somewhat lower bending and scattering losses than the Y-like modes. Moreover, for IDCF, anti-crossing occurs in the vicinity of a carrier frequency of 128 GHz for Y-like modes, which leads to high modal group velocity dispersion, and makes this mode undesirable. Therefore, in our experimental studies we favor X-like modes to the Y-like modes, especially for the IDCF.

## 4. Modal structure of the bent fiber

Here we present the theoretical studies of bending losses for the X-polarized and X-like modes of the three fibers [see Fig. 4(d)] for the two orthogonal fiber bending directions (parallel and perpendicular to the modal dominant polarization direction). As an example, the schematics of the two bending directions for the SSCF are shown in Fig. S2(a). There, the bend plane is either coincident with (left) or perpendicular to the X axis (right), which is also the polarization direction of the SSCF X-polarized mode. The bending radius ($R_b$) is defined as the distance from the fiber center to the bend axis. The simulations for the complex dispersion relations of the modes of a bend are carried out using COMSOL Multiphysics (axis-symmetric 2D model) using the same fiber cross-section as in Fig. 1, while considering the fiber material as lossless.

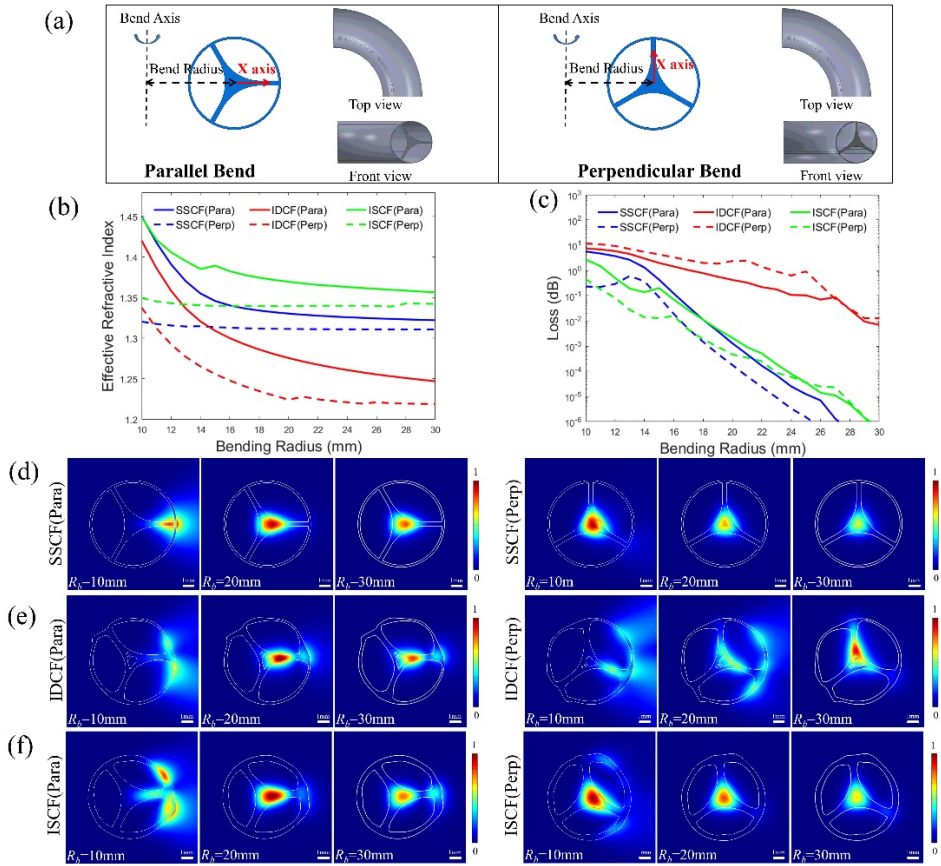

Fig. S2. (a) Schematics of the two types of 90° bends with 15 mm bending radius for SSCF. (b) Theoretical effective refractive indices and (c) bending losses per 90° bend (by power) of the X-polarized and X-like modes for the three fibers and the two types of bends as a function of the bending radius at 128 GHz. (d – f) Normalized electric field distributions of the X-polarized and X-like modes for the three fibers and the two types of bends as a function of the bending radius at 128 GHz.

In Figs. S2 (b),(c) we show complex dispersion relation of the X-polarized and X-like modes at 128 GHz propagating through two different bend types with bending radii $R_b$ ranging from 10 mm to 30 mm. These values are chosen to study degradation of the fiber performance under tight bending conditions that can be encountered in practical applications. We observe that, the real part of the modal effective refractive indices as well as modal losses can increase considerably for tighter bends, more so in the case of a IDCF that features weaker modal confinement in the core. This is due to increased modal presence in the solid core, bridge and cladding regions for tighter bends as clearly seen from the modal field distributions presented in Figs. S2(d – f).

Moreover, we also find that bends that are perpendicular to the modal polarization direction have smaller effect on the modal refractive index and losses than bends that are parallel to the modal polarization direction. This is clearly related to anisotropy in the fiber geometry. Particularly, for the perpendicular bends, the core boundary is almost perpendicular to the bend plane, and, as a consequence, modal fields are strongly confined in the fiber core by the high core/air refractive index contrast, thus preventing them from being pulled along the radial direction of the bend. In contrast, for parallel bends, the core boundary is parallel to the bend plane and nothing prevents the modal fields to be pulled along the bend radial direction into the bridge and cladding regions.

Finally, from Fig. S2 (c) we observe that for the SSCF and ISCF, bending losses smaller than ~ 0.1 dB per 90° bend for both bend types are achieved for bending radius larger than ~ 16 mm, while for IDCF one requires bending radius larger than ~ 27 mm. From this we conclude, that as far as bending losses are concerned, all the three fibers can readily tolerate tight bends with radius as small as 3 cm, resulting only in small loss increase. That said, we also note zero dispersion frequency of a bend mode will be somewhat shifted from that of a straight fiber. Also, additional losses are expected at the junction between a straight fiber and a bent section due to mismatch in the modal field distributions of the two waveguides. Therefore, a complete analysis of the degradation of the information capacity of a fiber link with bends is complex and deserves a separate study, while results of this section should be rather considered as an indication of the robustness of our fibers to bending perturbations.

## 5. CW THz spectroscopy system and transmission loss measurement

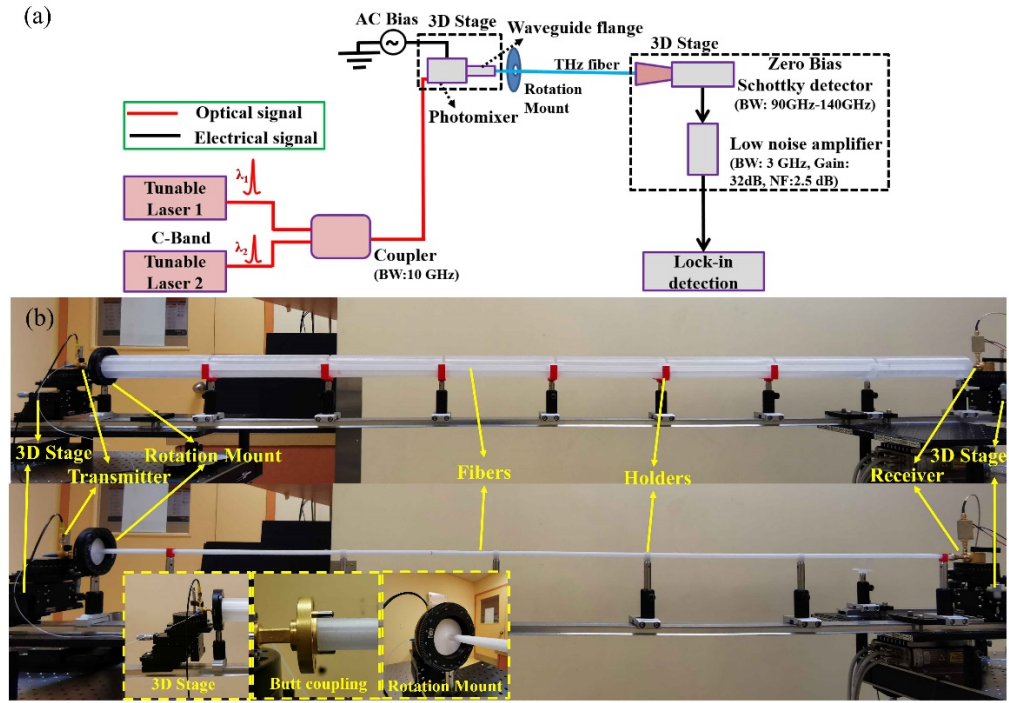

Fig. S3. (a) Schematic of CW THz spectroscopy system. (b) Experiment setup to measure the transmission losses for the SSCF (2 m) and ISCF (1.6 m). Insert: Enlarged view of the 3D stage, butt coupling of the ISCF with WR-6 rectangular flange, and the rotation mount.

The transmission loss measurements for all three fibers were characterized by using the CW THz spectroscopy system. The system schematic and experiment setup are shown in Fig. S3.

The CW THz spectroscopy system can be briefly described as follows: Two tunable lasers (TeraBeam) operating at judiciously mismatched wavelengths are used to optically drive the THz photomixer (emitter) for the difference frequency generation of THz waves. THz radiation from the photomixer (Model: IOD-PMD-14001from NTT Electronics Inc) is guided inside a WR-6 rectangular waveguide flange [Fig. 4(c)], which is butt-coupled to the fiber under study. On the receiving end, a 10.8 mm diameter horn antenna is used to collect THz waves, which are then detected and demodulated using a zero bias Schottky detector (Model: WR8.0 ZBD-F from Virginia diodes Inc). A high gain low noise amplifier (Model: SLNA-030-32-30-SMA from Fairview Microwave, Inc)is then used to amplify the received signal for further signal processing.

For the experiment setup, at the transmitter end, the THz photomixer is butt coupled to the WR-6 rectangular flange. At the receiver end, the THz signals are captured by the horn antenna. The polarization directions of the transmitter and receiver antenna were fixed to be horizontal as we planned for the preferential excitation of the X-polarized and X-like modes of the 3D printed fibers in the experiments. Both the transmitter and receiver were mounted on the 3D Stage (RBL13M from Thorlabs Inc) to finely optimize excitation and detection positions for the maximum detection signal. The fibers were mounted on a Rotation Mount (RSP2D from Thorlabs Inc) at the input side in order to rotate the whole fiber with a minimal step value of 2 degrees for optimizing the excitation inclination. For each fiber inclination angle, both the transmitter and receiver were rescanned to maximize the signal. Finally, the measurement was conducted after the optimal fiber inclination and the positions of transmitter and receiver were identified. In order to minimize the connector loss of SSCF, the fiber sections were mechanically joined using 3D-printed connectors shown in Fig. 2(e). Similarly, to minimize the effect of macro-bending in long fiber sections (IDCF and ISCF), multiple supports were fixed along the fiber length. After each measurement, position of the receiver was reoptimized for the maximal output power. The position of transmitter and the inclinations of all three fibers were fixed and unaltered throughout the loss measurements in order to maintain the optimal excitation arrangement (position and inclination).

## 6. Photonics-based THz communication system and BER measurement

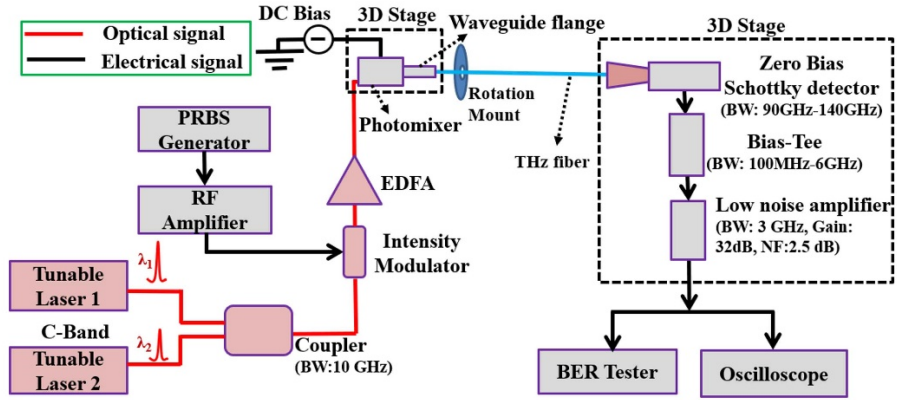

Fig. S4. Schematic of the photonics-based THz communication system.

The bit error rate (BER) measurement was conducted with the photonics-based THz communication system. The schematic of the THz communication system is shown in Fig. S4, the infrared optical signals are modulated using an external electro-optic modulator (Model: LN81S-FC and MX10A from Thorlabs, Inc) in the transmitter section whereas the received and demodulated baseband signals are recorded/analyzed using high-speed oscilloscope and BER tester (Model: MP2100B from Anritsu Corporation). The 3D printed fibers, the transmitter and receiver antenna were in a similar arrangement to the modal loss measurement (See Fig. S3(b)), meanwhile, the communication unit was enabled and a Bias-Tee was added before the low noise amplifier at the receiver side for conducting the BER measurement. A non-return-to-zero (NRZ) pseudo random bit sequence (PRBS) digital signal with a pattern length of $2^{31}-1$ was used as the baseband signal. The power of the THz transmitter antenna was set to -6.6 dBm (~218 μW), and the carrier frequency was chosen to be 128 GHz. The BER measurement was carried out by varying the bit rates from 1 Gbps to 6 Gbps. The target BER was set to $10^{-12}$ and the measurement duration is inversely proportional to the bit rate (1/(target BER · bit rate)). At each bit rate, the decision threshold was optimized in order to have a similar insertion (digital zero is mistaken as one) and omission errors (digital one is mistaken as zero) [3].

## 7. Experiment setup for mode field imaging

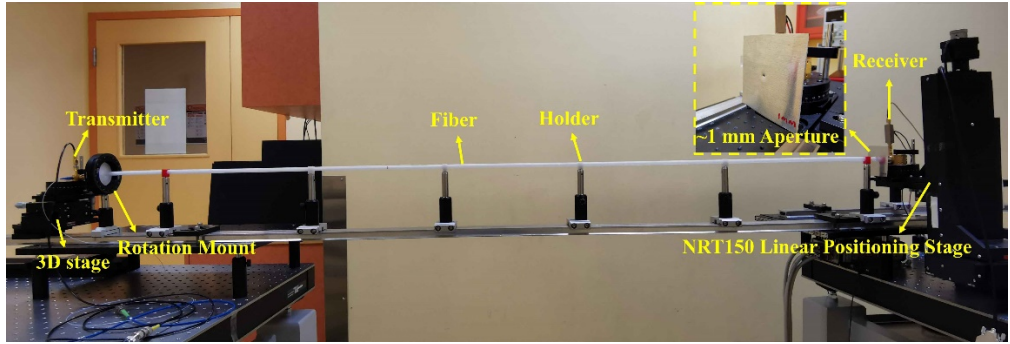

Fig. S5. Experiment setup for the near-field THz modal imaging of ISCF (1.6 m). Insert: Enlarged view of the subwavelength (~ 1 mm) aperture mounted on the horn antenna.

The near-field THz modal imaging for all three fibers was carried out by using the CW THz spectroscopy system as shown in Fig. S3(a). The 3D printed fibers, the transmitter and receiver antenna were in a similar arrangement to the modal loss measurement shown in Fig. S3(b), the difference is that the receiver was mounted on a 3-dimensional computer-controlled linear stage featuring on-axis accuracy of 2 μm (NRT150 from Thorlabs, Inc) and a sub-wavelength aperture of ~1 mm diameter (see the Inset of Fig. S5) was attached to the 10.8 mm diameter horn antenna of the detector to work as a near-field probe. The fiber excitation arrangement (position and rotation) was maintained the same as the transmission loss measurements. The modal field profile was acquired by raster scanning of the fiber end with spatial resolution of 0.3 mm, resulting in 41 x 41 images, each covering 12×12 mm$^2$ area of the fiber cross-section.